\documentclass{aa}  
\usepackage{graphicx}
\usepackage{txfonts}
\usepackage[colorlinks=true, linkcolor = blue, citecolor = blue]{hyperref}

\usepackage{amsmath}
\usepackage{multirow}
\usepackage{amsfonts}
\usepackage{comment}

\usepackage{xcolor}
\usepackage[normalem]{ulem} 

\def\ie{{\frenchspacing\it i.e.}~}
\def\eg{{\frenchspacing\it e.g.}~}

\begin{document}

   \title{A universal and physically motivated threshold for Hessian-based cosmic web identification: V-Web case}

   
   \author{Edward Olex \inst{1}
          \and Wojciech A. Hellwing \inst{2}
          \and Alexander Knebe \inst{1,3,4} 
          }

   \institute{Departamento de F\'isica Te\'{o}rica, M\'{o}dulo 15, Facultad de Ciencias, Universidad Aut\'{o}noma de Madrid, 28049 Madrid, Spain\\
              \email{edward.olex@uam.es}
        \and Center for Theoretical Physics, Polish Academy of Sciences, al. Lotników 32/46 Warsaw, Poland 
         \and Centro de Investigaci\'{o}n Avanzada en F\'isica Fundamental (CIAFF), Facultad de Ciencias, Universidad Aut\'{o}noma de Madrid, 28049 Madrid, Spain
         \and International Centre for Radio Astronomy Research, University of Western Australia, 35 Stirling Highway, Crawley, Western Australia 6009, Australia
        }
   \date{Received July 15, 2024; accepted October 16, 2024}

 
  \abstract
   {The study of large-scale structure can benefit from accurate and robust identification of the cosmic web. Having such classification can facilitate a more complete extraction of cosmological information encoded therein. This can help us to better map and understand galaxy-environment interactions. Classification methods like T-web and V-web, based on the Hessian matrix, are widely used to signal-out voids, sheets, filaments, and knots. However, these techniques depend on a threshold parameter which value is chosen without physical justification, usually relying on a user visual impression, thus limiting the universality of results.}
   {In this paper we focus on the V-web method. Our aim is to find a physical motivation for deriving an universal threshold that can be applied across different cosmic scales and epochs.}
   {V-web classify the large-scale structure using the eigenvalues of the velocity shear tensor. Using a set of gravity-only simulations we introduce a normalization that incorporates the standard deviation of the velocity divergence field, isolating the beyond-Gaussian evolution of cosmic web elements.}
   {In the Zeldovich's approximation, the probability presence of each cosmic web element remains constant at a threshold equal to 0. For the first time, we reveal that this behavior also holds in the non-linear regime for a normalized positive ``constant volume threshold'' that depends on both the redshift and the applied smoothing scale. The conservation of volume fractions is valid for the studied redshifts between 0 and 2, regardless of cosmic variance, and is most precise for intermediate smoothing scales around 3 Mpc/h. The properties of the cosmic web derived using this approach in the V-web align with expectations from other methods, including visual impressions. We provide a general fit formula to compute the constant volume threshold for any standard cosmological simulation, regardless of its specific properties.}
   {}


   \maketitle
   
%

\section{Introduction}

It is well understood that Dark Matter (DM) governs the large-scale distribution of the universe, because it is gravitationally dominant matter component (\cite{Planck2020}). Baryonic matter falls into the gravitational potential of DM and, for decades, it has been possible to trace its underlying distribution by means of galaxy surveys ranging from the first small samples from \cite{Huchra1983,Geller1989} to modern larger and deep catalogs like \cite{Euclid2022}. The formation and evolution of this inhomogeneous matter distribution is dominated by the gravitational instability mechanism. The process can be treated analytically by linear and quasi-linear approximations that predict the early stages of DM evolution \citet{Zeldovich1970,Peebles1980}. In the subsequent non-linear regime, the DM forms halos by gravitational collapse, where the observed galaxies are located \citep{White1978}, and these in turn are immersed in a larger web-like structure called the Cosmic Web (CW, \citet{Bond1996}). Both the properties and evolution of CW can be studied directly in cosmological simulations and are key to understanding the connection between the DM haloes and their environment \citep[\eg][]{Hahn2007,Maccio2007,Cautun2014,Alpaslan2014,Hellwing2021,MHunde2024}. In addition, the structure around the galaxies affects their star formation \citep[\eg][]{Peng2012,Darvish2017,AragonCalvo2019,Jaber2024}, spin \citep[\eg][]{Zhang2015,Pahwa2016,Punya2019}, accretion of satellites \citep[\eg][]{Knebe2004,2019AstBu..74..111K}, morphology \citep[\eg][]{Dressler1980}, and origin \citep[\eg][]{Codis2012}. Moreover, the CW can shed light on the standard cosmological model \citep{Codis2018,Bonnaire2022} and alternatives to General Relativity \citep[\eg][]{Falck2015,Dome2023,SGupta2024}.

The classification of CW elements can be carried out analytically in the linear regime using the \cite{Zeldovich1970} approximation, which divides the structure into Voids, Sheets, Filaments and Knots depending on the value of the eigenvalues of the linear deformation tensor. The classification is performed taking into account the sign of the three available eigenvalues and interpreting it as a collapse (``$+$'') or expansion (``$-$'') in the direction of the corresponding eigenvector. Therefore, the element type can be assigned using only the density field, classifying each region of space into Void, Sheet, Filament or Knot. This strategy has also been used to study the non-linear regime in cosmological simulations using other tensors such as the Hessian of the gravitational potential aka tidal field \citep[T-web][]{Hahn2007, Calvo2007, Romero2009}, and the velocity shear tensor (V-web, \cite{Hoffman2012}). Other strategies are based on using the shear of the Lagrangian displacement field (DIVA, \cite{Lavaux2010}) or on the number of orthogonal axes in the Lagrangian phase space sheet (ORIGAMI, \cite{Falck2012}). The density field is also widely used in several studies based on segmentation (Spineweb, \cite{Calvo2010}), finding its topological structures using Morse theory (\cite{Colombi2000}), or applying adaptive scaling methods to search for filaments (NEXUS, \cite{Cautun2012}). See \cite{Libeskind2017} for a summary and comparison of the CW identification methods.

The benefit of CW studies via the means of the T-web and V-web (and any of their variations based on Hessian matrix eigenvalues) relies on its easy and straightforward application. The additional advantages consists of their generalization in various types of cosmological simulations, the analytical formulation for T-web in the linear and quasi-linear regime (\cite{Emma2023}), and the possibility of apply both methods to observations, via the projected galaxy distribution for T-web (\citet{Alonso2016}) and radial peculiar velocities for the V-web (\citet{Pomarede2017}). T-web and V-web are equivalent in the linear regime because, at the first perturbative order, the velocity of matter is proportional to its gravitational field.

Although Zeldovich's approach suggests separating collapse and expansion using the sign of the eigenvalues, subsequent studies in the non-linear regime confirm better agreement with the distribution of the fields using a positive number between 0 and 1 to separate both situations. Thus the 'threshold problem' is born: the results of T-web and V-web depend on a free parameter $\lambda_{th}$ called threshold, which depends (in most of the cases) on the visual impression of each author. In the literature several values are used for the T-web, among which $\lambda_{th}=0$ is used by \cite{Hahn2007} as in the linear case, $\lambda_{th}=0.01$ by \cite{Weiguang2017}, $\lambda_{th}=0.2$ by \cite{Libeskind2014}, or $\lambda_{th}\in[0.2,0.4]$ by \cite{Romero2009} among others. For the V-web, the range $\lambda_{th}\in[0,0.6]$ has been used by \cite{Pfeifer2022}, and $\lambda_{th}=0.44$ by \cite{Hoffman2012} and \cite{Libeskind2014}, among others. In many cases the authors have to study their results based on various thresholds since the conclusions may change with respect to this free parameter. This fact has hindered the use of eigenvalue-based methods to study the time evolution of CW as well as its generalization across scales, since for each situation a different threshold would have to be used to visually adjust the matter distribution. It has become an important goal to find a physical motivation to justify a concrete threshold value in T-web and V-web. Not only to regularize the results of these methods, but also to build a complete non-linear physical model of the CW without the need for free parameters. 

For this purpose, in this paper we have focused mainly on V-web, although we briefly comment on the possible extension to T-web in the final discussion. We study a wide range of possible $\lambda_{th}$ values in a suite of gravity-only $\Lambda$CDM simulations with different initial conditions. In each simulation we make $\lambda_{th}$ proportional to the variance of the velocity field divergence. This normalization allows for a direct comparison of the volume fractions (or probabilities) of each CW element across different redshifts. When analyzing the time evolution instead of considering each redshift independently, a single threshold value $\lambda_{CV}$ consistently arises across all simulations, at which the fractions of all CW elements remain approximately constant. However, although the volume fraction of each element is conserved, this is not the case for its mass fraction, which shows a time evolution in line with what is expected \citep[see \eg][]{Cautun2014, Zhu2017}. To give a more universal solution that would be applicable to any independent simulations, we provide a fit to obtain $\lambda_{CV}$ depending on the scale and redshift. Testing the fit with an off-set simulation run (with a different baseline cosmology), recovers  the expected result and thus confirming the generality of our solution.

This fact allows us to analyze the characteristics of CW and its evolution with V-web that does not require the choice of an arbitrary $\lambda_{th}$ based on visual impression. Moreover, it sheds light on possible new physical behavior in the large-scale distribution of DM related to the conservation of CW volume fractions. The subsequent work will be the construction of an analytical framework based on the quasi-linear approximation from which to explain the conservation of probability in the V-web classification and to adjust the presented results for simulations. Both perspectives will allow us to get closer to a general theory of CW not yet constructed, and to understand in more depth the relationships between dark matter halos and large-scale structure.

This paper is organized as follows: in section \ref{sec-hessian_classification} we summarize the basis of Hessian methods for CW classification and we describe the set of cosmological simulations used. In section \ref{sec-evolution} we detail how conservation of volume fractions arises naturally from a suitable eigenvalue normalization, and in section \ref{sec-VF_and_MF} we show its consequences in general CW statistics. In section \ref{sec-application} we give indications on the application of the constant volume threshold to generic simulations and section \ref{sec-conclusions} is a brief discussion of the results and some general conclusions.


\section{Hessian-based Cosmic Web classification}
\label{sec-hessian_classification}

By constructing a grid in a gravity-only simulation snapshot, discrete fields located along the volume such as density contrast $\delta(\vec{r})$, gravitational potential $\phi(\vec{r})$ and velocity $\vec{v}(\vec{r})$ can be obtained. T-web \citep{Forero2009} uses the Hessian of $\phi(\vec{r})$:

\begin{equation}
    T_{ij} = \frac{\partial ^2 \phi}{\partial{r_i}\partial{r_j}}    
\label{eq-hessian}
\end{equation}

to classify each grid cell located in a given position $\vec{r}$ into $4$ types according to how many eigenvalues of the $T_{ij}$ tensor are above a certain threshold $\lambda_{th}$. These eigenvalues are labeled according to the order $\lambda_{1} \leq \lambda_{2} \leq \lambda_{3}$. If at $\vec{r}$ no eigenvalue is greater than $\lambda_{th}$, the cell volume is considered to be a void. If only the largest one, $\lambda_1$, is above the threshold, the point is classified as a sheet, for first two a filament and for all three a knot. Intuitively, if $\lambda_{i} < \lambda_{th}$ along a certain direction, it indicates expansion in that direction. Conversely, if $\lambda_{i} > \lambda_{th}$, it signifies compression along that axis. Expansion along all three dimensions implies extensive voids, while in contrast compression along all three directions signify dense knots. The intermediate situations depend on whether the expansion is in one dimension (filaments) or in two (sheets). Table~\ref{tab-table1} shows a summary of this classification. 

\begin{table}
\caption{Classification of the cosmic web by eigenvalues used in the T-web and V-web methods. The condition $\lambda_{i} - \lambda_{th} \leq 0$ is denoted by ``$-$'', and the condition $\lambda_{i} - \lambda_{th} > 0$ is denoted by ``$+$''.}             
\label{tab-table1}      
\centering                          
\begin{tabular}{c c c c}        
Element & $\lambda_{1}$ & $\lambda_{2}$ & $\lambda_{3}$ \\    
\hline                        
   Void & $-$ & $-$ & $-$ \\      
   Sheet & $+$ & $-$    & $-$ \\
   Filament & $+$ & $+$     & $-$ \\
   Knot & $+$ & $+$    & $+$ \\
\hline                                   
\end{tabular}
\end{table}

Alternatively, V-web \citep{Hoffman2012} use the eigenvalues of the velocity shear tensor defined as:

\begin{equation}
    \Sigma_{ij} = -\frac{1}{2H(z)} \left(\frac{\partial{v_j}}{\partial{r_i}}+\frac{\partial{v_i}}{\partial{r_j}}\right)   
\label{shear}
\end{equation}

where the Hubble parameter $H(z)$ scales the tensor for each redshift and makes it dimensionless (\cite{Libeskind2014}). The ``$-$'' sign included in the definition (\ref{shear}) makes the collapse and expansion conditions the same as in Table~\ref{tab-table1}. For each grid cell we can further take the trace of the velocity shear tensor $Tr(\Sigma)$ resulting in the velocity divergence field $\theta(z)$:

\begin{equation}
    \theta(z) \equiv Tr(\Sigma) = -\frac{\nabla \cdot \vec{v}}{H(z)} 
\label{divergence}
\end{equation}

Prior to diagonalization, a smoothing is performed to mitigate the effects of the grid discretization. For this purpose we use a Gaussian kernel of the form:

\begin{equation}
    W_{G}(k, R_{s}) = \exp \left( -\frac{1}{2} k^{2}R_{s}^{2} \right)
\label{kernel}
\end{equation}

where $R_s$ is a comoving smoothing scale in Mpc/h units also denominated smoothing length.This distance has to be greater than twice the grid spacing used, since at smaller distances the convolution would not affect contiguous cells. It should also not exceed half the side of the simulated box. 

We can calculate the standard deviation of the divergence field $\langle \theta^{2} \rangle^{1/2}=\sigma_{\theta}$, being its value lower for early times with a homogeneous velocity distribution and higher for late times involving larger non-Gaussianities. This relationship follows from $\nabla \cdot \vec{v} \propto -\delta$ in the linear regime \citep{Peebles1980,Chodorowski1997}. Similarly $\sigma_{\theta}$ is smaller at higher smoothing lengths $R_{s}$ since in linear regime $\sigma_{\theta} \propto \sigma_{\delta}$ and:

\begin{equation}
    \sigma_{\delta}^2(z) = {1\over{2\pi^2}}\int dk\, k^2\, P(k) W_{G}(k, R_s)^2\,
\label{eq-linearstd}
\end{equation}

where $P(k)$ is the matter power spectrum. If we assume that the contrast density field $\delta$ is Gaussian, then $\theta$ is also Gaussian and therefore we can compute the joint probability of the eigenvalues of $\Sigma_{ij}$ using Doroshkevich's formula (\cite{Doroshkevich1970}):

\begin{equation}
\begin{split}
p(\lambda_{1},\lambda_{2},\lambda_{3}) = \frac{3375}{8\sqrt{5}\pi\sigma_{\theta}^{6}}\exp\bigg[-\frac{3}{\sigma_{\theta}^{2}}\left(\sum_{i=1}^{3}\lambda_{i}\right)^2 \\
 +\frac{15}{2\sigma_{\theta}^{2}}\sum_{i>j}\lambda_{i}\lambda_{j}\bigg]\, \vert\Pi_{i >j}(\lambda_{i}-\lambda_{j})\vert
\end{split}
\label{eq-doro}
\end{equation}

Therefore, the probability $P^{\rm E}$ of having an element type E in the V-web classification in the Gaussian case depend on the joint probability given by Eq.~(\ref{eq-doro}) and the condition determined for each element in Table~\ref{tab-table1} (\cite{Emma2023}):

\begin{equation}
    P^{\rm E}(\lambda_{th}) = \int d\lambda_{1}d\lambda_{2}d\lambda_{3}\, p(\lambda_{1},\lambda_{2},\lambda_{3})\,C^{\rm E}(\lambda_{1},\lambda_{2},\lambda_{3},\lambda_{th})
\label{eq-pe}
\end{equation}

where $C^{\rm E}$ is the function that contains the Boolean condition of each element, being 1 if it is satisfied or 0 if not. The integral $P^{\rm E}$ depend not only on the threshold, but also on parameters such as redshift, smoothing length, and cosmology due to the presence of $\sigma_{\theta}$ in Eq.~(\ref{eq-doro}). The equivalent of the probability of an environment E in cosmological simulations is the volume fraction $f_{v}^{\rm E}(\lambda_{th})$, which is obtained by summing the volume occupied by element E and dividing by the total volume of the box. Under periodic boundary conditions $P^{\rm E} \simeq f_{v}^{\rm E}$. To investigate the non-Gaussian regime where this equality no longer applies, we generate a set of cosmological simulations designed to accurately reproduce the CW at low redshift.

\paragraph*{\textbf{Simulations}} ~\\
We employ the {\it COmoving Lagrangian Accelerator} (COLA) method \citep{Tassev2013} to obtain\
our set of N-body simulations. The parallel implementation of the COLA algorithm (called PI-COLA, see \citep{Howlett2015}) allows to run large simulations at a reduced computational cost, with the trade-off of limited spatial and temporal resolution. Specifically, we use a publicly available optimized branch of the PI-COLA family, the MG-COLA, developed by \citet{Winther2017}.
COLA achieves the significant speed-up mostly at the expense of the accuracy of DM halo properties. 
For example, the method is  known to bias weakly the resulting halo velocities. This bias is small (up to $\sim5\%$ for our case) and concerns mostly the small-scale halo velocity field (see more in Ref.~\citep{Munari2017}).
However, the COLA simulations convergence to the full N-body results is very good for the large-scale smooth density and velocity fields for the scales and epoch we consider in this paper. 
We use a set of $10$ gravity-only simulations in MG-COLA with a box size of $L = 250$ Mpc/h and with $N_{p} = 512^3$ particles each representing $m_p=8.6\times 10^{9} M_{\sun}/h$ matter element, under periodical boundary conditions. All simulations have different initial phases. To ensure sufficient accuracy of the COLA method at small scales, we use a total of 15 time steps from the initial conditions (3 steps linearly distributed between the redshifts $z=0,0.3,0.5,1,2,10$). The cosmological parameters used are: $\Omega_m = 0.267$, $\Omega_b = 0.049$, $h=0.71$, $\sigma_8 = 0.8$ and $n=0.966$.

\section{Evolution of the cosmic web}
\label{sec-evolution}

The typical way in which V-web and T-web are used is to perform a static classification at a given redshift, using a preselected threshold $\lambda_{th}$ chosen by visual inspection. This approach prohibits both of the methods to be used for a consistent analysis of the CW evolution. We show that accounting for the time-evolution of the large-scale cosmic fields, including their associated Hessian eigenvalues, enables the determination of a unique, time-dependent mapping for the optimal (and natural) V-web threshold $\lambda_{CV}$. We find that this mapping is robust and universal, and can be used for consistent V-web components identification across redshift, simulations and cosmologies.

\subsection{The threshold problem}

\begin{figure*}
\centering
\includegraphics[width=17.5cm]{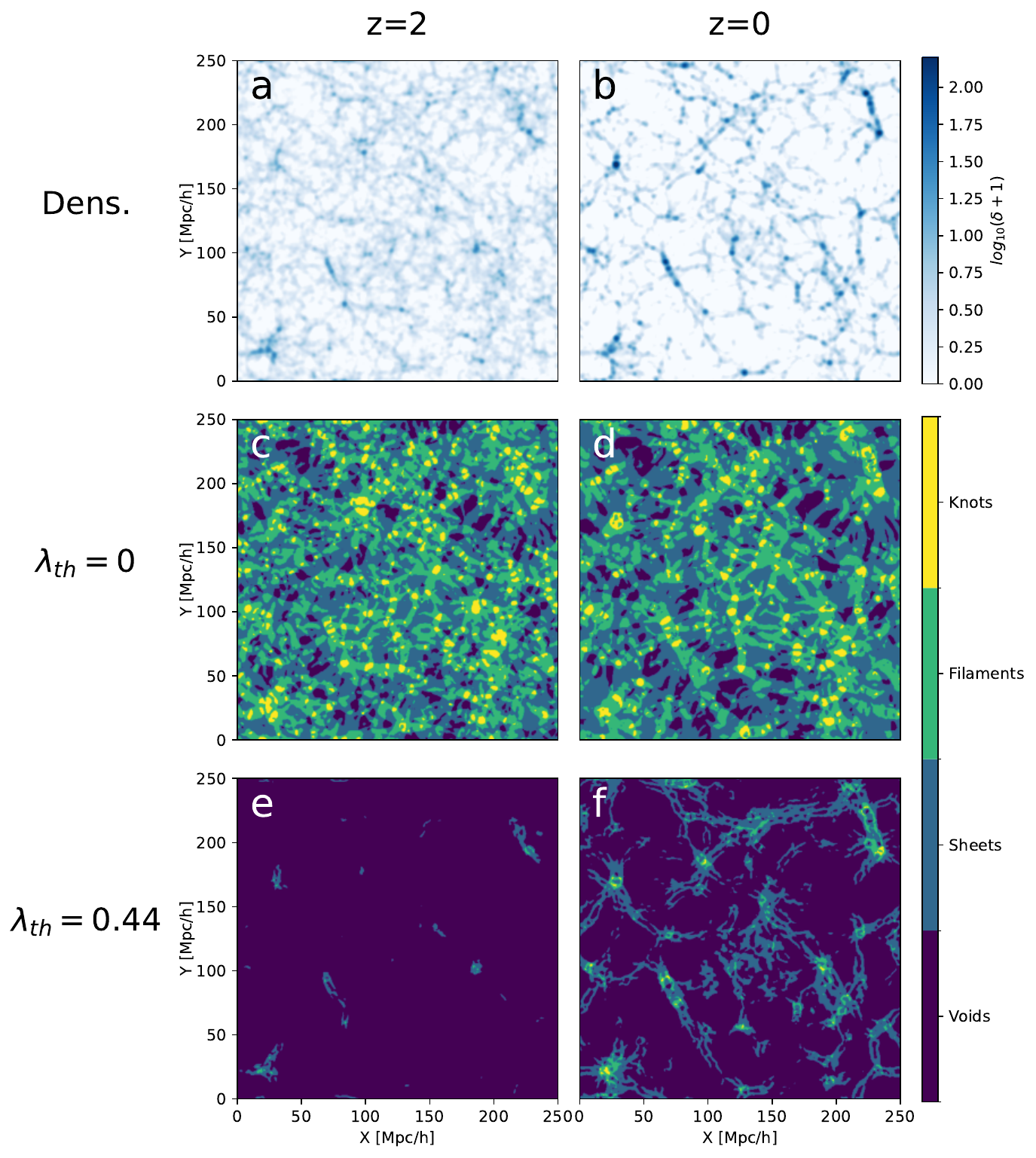}
\caption{Density and V-web classification of a gravity-only $\Lambda$CDM simulation. The cube is divided into a $N_{g}=512^3$ cells grid on which both the density and the eigenvalues of the shear velocity tensor have been calculated, applying a smoothing of $R_{s} = 1$ Mpc/h. Panels a and b show the logarithm of the dark matter density at redshifts 2 and 0 respectively (in a slice that has a thickness of $\sim 1.5$ Mpc/h). Panels c and d contain, for the previous redshifts, the classification into CW elements performed by the V-web method for an $\lambda_{th}=0$ (in a middle slice that is one grid cell thick). Panels e and f contain the same as the previous two but for the $\lambda_{th}=0.44$ proposed by \cite{Hoffman2012}.}
          \label{one_page_first}%
\end{figure*}

The example of an N-body simulation of our set in which V-web was applied is shown in Fig.~\ref{one_page_first}. Using a smoothing length of $R_{s} = 1$ Mpc/h, the density and velocity fields are computed on a $N_{g}=512^3$ grid, as well as the classification of the CW elements according to the value of the eigenvalues of Eq.~(\ref{shear}). Panels \hyperref[one_page_first]{a} and \hyperref[one_page_first]{b} contain the logarithm of the dark matter density distribution $log_{10}(\delta + 1)$ computed on a $\sim 1.5$ Mpc/h thick slice (or three times the grid spacing), for $z=2$ and $z=0$ respectively. For both redshifts, it is possible to distinguish zones with higher matter density, which usually end up forming galaxy clusters, and zones with lower density that tend to empty as the matter flows towards the regions with higher concentrations. It can be noted that the CW must be mostly occupied by voids, with the denser knot-like zones bound by thin sheets and filaments.

The classification made by the V-web in the slice formed by the cells in the middle is shown in the rest of the panels. For \hyperref[one_page_first]{c} and \hyperref[one_page_first]{d}, where we have applied a threshold $\lambda_{th}=0$, there is an excess of structure that leaves voids as a minority element compared to filaments and sheets. This feature contradicts the main observations that predict a universe mostly full of voids (see \cite{Gregory1978}, \cite{Joeveer1978} and \cite{Tully1978} for first detections), making the use of $\lambda_{th}=0$ unfavorable. In panels \hyperref[one_page_first]{e} and \hyperref[one_page_first]{f} we have applied $\lambda_{th}=0.44$ motivated by \cite{Hoffman2012}, which significantly improves the closeness to the visual impression at $z=0$. However, a constant threshold cannot account for the non-linear evolution of the eigenvalues $\lambda_{i}$, as it neglects the fact that these values are generally closer to 0 in a younger universe with more homogeneous density and velocity fields. Consequently, it results in an overly high threshold for $z=2$, leading to a scenario where almost the entire simulated space is classified as a void, thereby losing the structural details.

Table~\ref{tab-table2} contains the volume fractions computed for the simulation shown in Fig.~\ref{one_page_first}. The scarce presence of voids for the case $\lambda_{th}=0$ occurs because they occupy less than $15\%$ of the total volume at all times, as opposed to filaments and sheets that occupy more than $80\%$ of the space. In contrast, when $\lambda_{th}=0.44$ is applied at $z=2$, most of the $\lambda_{i}$ values fall below $\lambda_{th}$, resulting in over 99$\%$ of void fraction. The threshold problem extends not only to the redshift evolution, but also to the smoothing scales used, having to adjust the $\lambda_{th}$ value when the scale is changed. This fact prevents the use of V-web (and T-web) in an automated and self-consistent way.

\subsection{The dependence of the volume fractions evolution on the threshold}
\label{sec-3.2}

Since the volume fractions can also be understood in the probabilistic sense, in Table~\ref{tab-table2} the values of $P^{\rm E}$ defined in (\ref{eq-pe}) for the selected thresholds are shown in parentheses. Since Eq.~(\ref{eq-doro}) only works for Gaussian fields, the linear result can only give a slight approximation of the values obtained for the simulation. Note that for the value suggested by Zeldovich's approximation $\lambda_{th}=0$ the linear volume fractions do not change along the redshift, this being the only value at which this conservation is satisfied.

\begin{table}
\caption{Volume fractions of the CW elements calculated with V-web for the examples shown in Fig.~\ref{one_page_first} using $R_{s} = 1$ Mpc/h. The $P^{\rm E}$ values obtained from Eq.~(\ref{eq-pe}) using $\sigma_{\theta}$ from the simulation are written in parentheses.}             
\label{tab-table2}      
\centering                          
\begin{tabular}{c c c}        
   &  $z=2$ &  $z=0$  \\
\hline \\ [-2ex]
   $\lambda_{th}=0.00$ &  $f_{v}\,(P)$ &  $f_{v}\,(P)$  \\
  Void & 0.11 (0.080) & 0.14 (0.080)  \\      
     Sheet & 0.46 (0.42) & 0.49 (0.42)  \\
     Filament & 0.37 (0.42) & 0.31 (0.42) \\
     Knot & 0.061 (0.080) & 0.045 (0.080)  \\
\hline \\ [-2ex]               
 $\lambda_{th}=0.44$ &  $f_{v}\,(P)$ &  $f_{v}\,(P)$  \\
  Void & 0.99 (1.00) & 0.77 (0.71)  \\      
     Sheet & 0.0099 (4.3$\times10^{-5}$) & 0.20 (0.27)  \\
     Filament & 3.8$\times10^{-4}$ (4.6$\times10^{-9}$) & 0.029 (0.021) \\
     Knot & 4.3$\times10^{-6}$ (2.3$\times10^{-13}$) & 0.0012 (0.0003)  \\
\hline    
\end{tabular}
\end{table}

\begin{figure*}[h!]
\centering
\includegraphics[width=15cm]{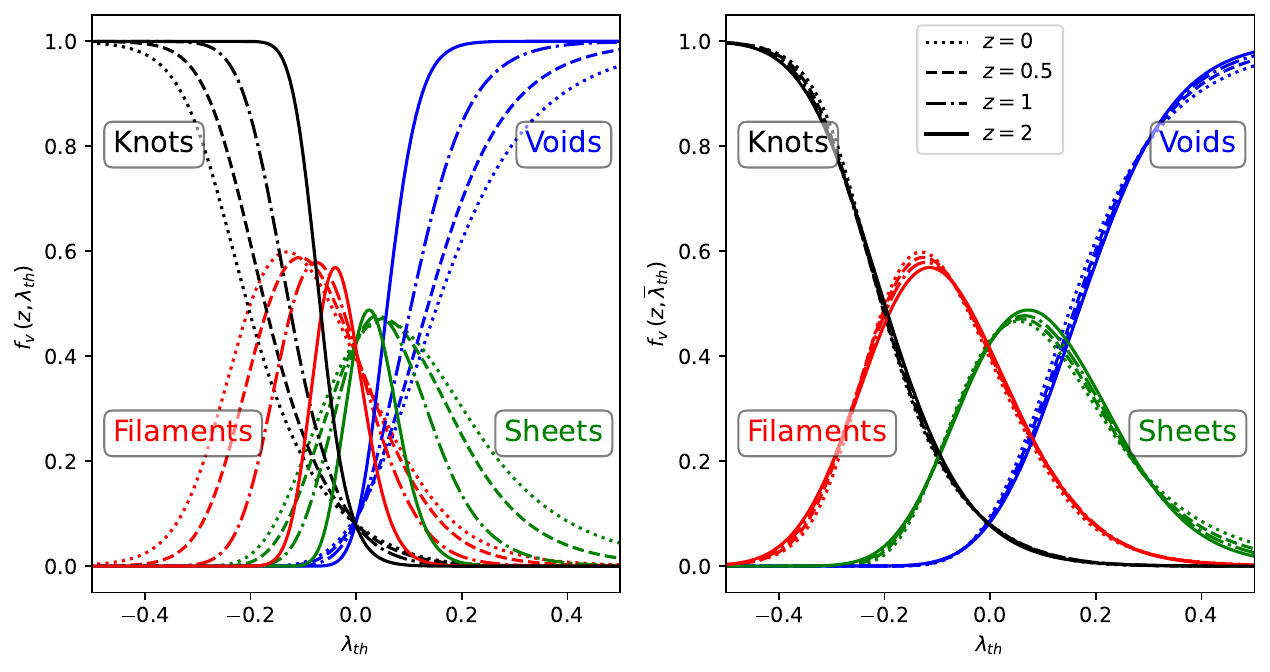}
\caption{Volume fractions of CW elements obtained with V-web from a simulation shown in Fig.~\ref{one_page_first}. The left panel show the results using the same threshold for all redshifts, while in the right panels Eq.~(\ref{eq-norm}) has been used to normalize $\lambda_{th}$. The applied smoothing length is $R_{s}=3$ Mpc/h. }
          \label{VandM_frac}
\end{figure*}

If we increase a little the value of $\lambda_{th}$ above $0$, it is more likely that $\lambda_{2}$ and $\lambda_{3}$ are smaller than the threshold so we expect an increase in the volume fraction of both the sheets and voids. If we continue to increase $\lambda_{th}$, there is a certain moment when $\lambda_{th}$ is above all $\lambda_{i}$ so the fraction of voids is greater than that of the sheets. Finally, for a $\lambda_{th}>\lambda_1\geq\lambda_2\geq\lambda_3$ case the voids occupy all of the space. The left panel of Fig.~\ref{VandM_frac} shows the evolution of the V-web volume fraction $f_{v}(z,\lambda_{th})$ at $R_{s}=3$ Mpc/h for the simulation presented in the previous figure. For $\lambda_{th} > 0$ the behavior described for sheets and voids can be observed. In the case $\lambda_{th}<0$ a similar behavior occurs for filaments and knots.

It can be seen in the left panel of Fig.~\ref{VandM_frac} that the solid lines representing $z=2$ are closer to $\lambda_{th}=0$ than the other redshifts because the distribution of eigenvalues is more restricted for a younger (\ie more homogeneous) universe. As the non-linear evolution progresses, the variance of the velocity field grows, and the eigenvalues can take a larger range of values. This causes the volume fractions to stretch at lower redshifts, extending their presence to threshold values farther from zero. We can normalize $\lambda_{th}$ with the variance of the velocity divergence $\sigma_{\theta}(z)$ to partially account for this effect. Therefore, we studied the volume fractions using a normalized $\overline{\lambda}_{th}$ defined as:
\begin{equation}
   \sigma_\theta^0\equiv\sigma_\theta(z=0)\;\quad, \overline{\lambda}_{th}(z) = \frac{\sigma_{\theta}(z)}{\sigma_{\theta}^0}\lambda_{th}   
\label{eq-norm}
\end{equation}
This normalization allows us to cancel the linear evolution of the CW because using $\overline{\lambda}_{th}$ as a threshold in the V-web is equivalent to multiplying $\lambda_{i}$ by the factor $\sigma_{\theta}(z)/\sigma_{\theta}^0$ in the conditions of Table~\ref{tab-table1}.
The above transformation applied to Doroshkevich's formula (\ref{eq-doro}) and to the differentials of Eq.~(\ref{eq-pe}) cancels all $\sigma_{\theta}(z)$ terms of the $P^{\rm E}$ function and replaces them with $\sigma_{\theta}^0$, which implies the loss of time dependence and
\begin{equation}
    P^{\rm E}(z,\overline{\lambda}_{th}) = P^E(z=0,\lambda_{th})    
\label{normpe}
\end{equation}
for a Gaussian case, so we can expect similar (but not identical) behavior for the volume fractions $f_{v}$ of the CW fields extracted from cosmological simulations. 

In the following, when we consider $z>0$ snapshots, we apply $\overline{\lambda}_{th}(z)$ as a threshold in the V-web to extract volume fractions or other quantities. To compare properties obtained using different $\overline{\lambda}_{th}(z)$ , we use $\lambda_{th}$ common to all redshifts by Eq.~(\ref{eq-norm}). The right panel of Fig.~\ref{VandM_frac} contains the volume fractions for the threshold values after normalization (\ref{eq-norm}). The x-axis represents $\lambda_{th}$; therefore, for each snapshot at redshift $z$, the volume fraction $f_{v}(z,\overline{\lambda}_{th})$ is obtained using the threshold $\overline{\lambda}_{th}(z)$ in the V-web. This strategy reduces the difference between the redshifts and brings all the volume fractions in each element closer to those obtained at $z=0$. However, there are still small differences depending on $z$ and $\lambda_{th}$ stemming from non-linear gravitational evolution that drives the cosmic fields away from a Gaussian distribution for late times and/or small scales. 

\begin{figure}
\centering
\includegraphics[width=\hsize]{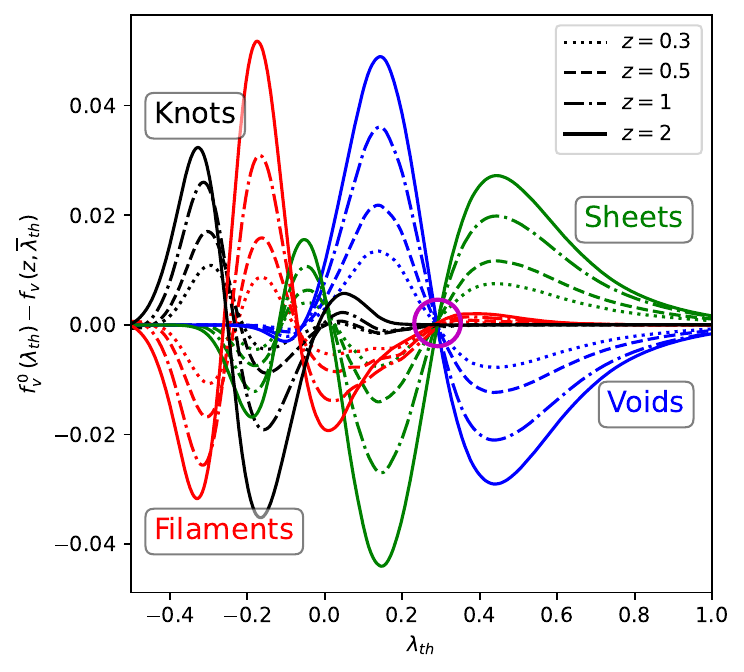}
    \caption{Evolution of the difference in volume fractions after $\lambda_{th}$ normalization defined in Eq.~(\ref{eq-norm}). Each line is the volume fraction at $z=0$ minus the volume fraction at other redshifts (differentiated with the type of pattern), as a function of $\lambda_{th}$. The color code, simulation and $R_{S}$ value is the same as the one used in Fig.~\ref{VandM_frac}. The magenta circle marks the constant volume threshold $\lambda_{CV}$ in which the volume fractions of each element become approximately constant over time.}
     \label{cRs3_Difvsth}
\end{figure}

This can be better seen in Fig.~\ref{cRs3_Difvsth} where we show the difference between the volume fractions at the present time $f_{v}^0(\lambda_{th}) \equiv f_{v}(z=0,\lambda_{th} )$ minus the volume fractions at the four other redshifts, for the same initial conditions and smoothing length as in Fig.~\ref{VandM_frac}. At the extreme $\lambda_{th}$, the differences tend to be zero for all elements because, at all redshifts, there is a dominance of knots at $\lambda_{th}<0$, and the dominance of voids at $\lambda_{th}>0$. However, the intermediate values contain information on how the volume fractions vary with time in mixed regimes with all CW elements present in the universe, so we restrict the analysis to thresholds where no single element exceeds a volume fraction of 0.99.
 
 In Fig.~\ref{cRs3_Difvsth} we note that there is a special $\lambda_{th}$ which, despite being intermediate to the two extreme situations, cancels out the volume differences at this value for all redshifts and CW elements. Indicated by the magenta circle, the volume fraction differences of all CW elements reach a minimum in this "constant volume" threshold $\lambda_{CV}$. This means that, if we choose as threshold the related value $\overline{\lambda}_{CV}(z)$ for the V-web, we always have approximately the same volume fractions for $z$ and $z=0$. This effect occurs because the curve for voids, filaments, and sheets changes sign at this $\lambda_{CV}$. Regarding the knot line, it appears that it tends to very low values from $\lambda_{th}$ less than $\lambda_{CV}$ to the extreme dominated by voids. The same behavior was found for $R_{s}$ between $1$ and $6.5$ Mpc/h, but with different values of $\lambda_{CV}$ depending on the case.
 
 Note that $\lambda_{CV}$ may be redshift-dependent because its value differ slightly depending on the $z$ at which the volume fractions are computed. To locate $\lambda_{CV}$ in a multitude of different simulations and redshifts, we define the function $S$ as:
 \begin{equation}
    S(z,\lambda_{th}) = \sqrt{ \sum_{\rm E}^{}|f_{v}^{\,0,\rm E}(\lambda_{th})-f_{v}^{\rm E}(z,\overline{\lambda}_{th})|^2}\,\,,
\label{S}
\end{equation}
thus using the sum of the differences for all CW elements.  

\begin{figure}
\centering
\includegraphics[width=\hsize]{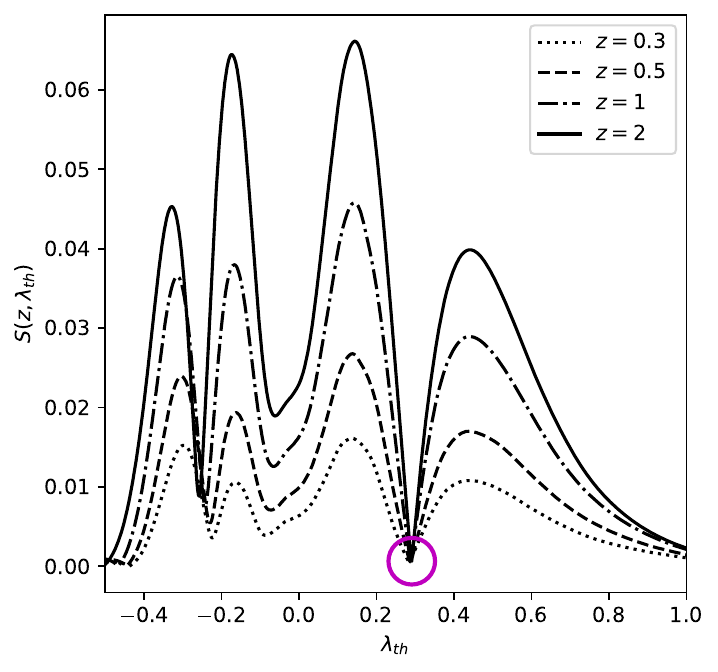}
  \caption{The function S defined in Eq.~(\ref{S}) calculated for the simulation of the previous figures ($R_{s} = 3$ Mpc/h). The magenta circle marks $S(\lambda_{CV})$.}
     \label{cRs3_Svsth}
\end{figure}

Function $S$ is shown in Fig.~\ref{cRs3_Svsth} for different redshifts. The threshold $\lambda_{CV}$, marked with a magenta circle, corresponds to a local minimum of $S$ in the intermediate zone between extremes. This minimum also appears for the other redshifts in a similar threshold. Given any gravity-only simulation, it is possible to find an approximate value of $\lambda_{CV}$ by first calculating its $f_{v}$ with threshold $\overline{\lambda}_{th}(z)$ as a function of $\lambda_{th}$ for different redshifts, and then computing the local minimum of $S$ between 0 and 1. Once $\lambda_{CV}$ is found for a given $z$, it is possible to perform a V-web analysis by applying $\lambda_{CV}$ to the $z=0$ snapshot and $\overline{\lambda}_{CV}(z)$ for the snapshot with a redshift $z$. 

We have used the same methodology for $10$ gravity-only $\Lambda$CDM simulations (described in section \ref{sec-hessian_classification}), where the V-web was applied on a grid of $512^{3}$ cells. The only difference between the simulations is the random seed used to pick the phases in the initial conditions. In all of them we have found an conservation of the volume fractions at a specific $\lambda_{CV}$, always similar to the behavior of Fig.~\ref{cRs3_Difvsth} in which the differences of each CW element tend to $0$ at a similar threshold value.

\subsection{Constant volume threshold $\lambda_{CV}$}

\begin{figure}
\centering
\includegraphics[width=\hsize]{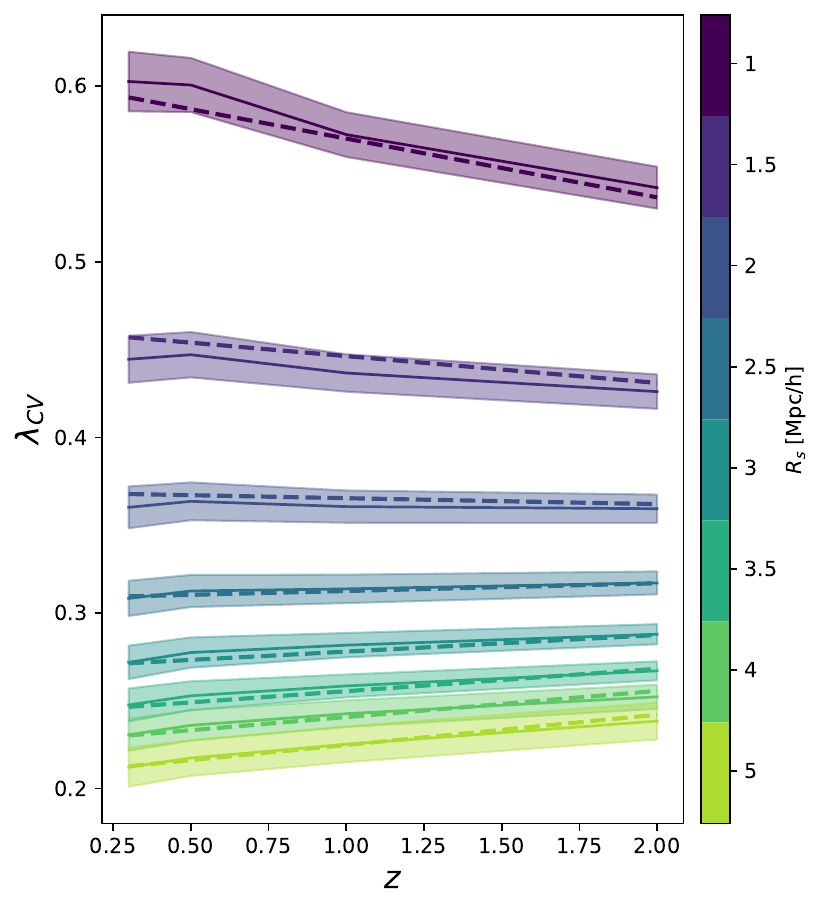}
  \caption{The solid lines shows the mean of $\lambda_{CV}$ computed using the minimum of the function $S$ in each simulation of our set. The dashed lines correspond to the fit defined in Eq.~(\ref{eq-fit2}) (see section \ref{sec-fit}). The different colors correspond to the smoothing length used in the analysis, with the shading indicating the $\sigma$ region of the set.
          }
     \label{thresh_z}
\end{figure}

\begin{figure}
\centering
\includegraphics[width=\hsize]{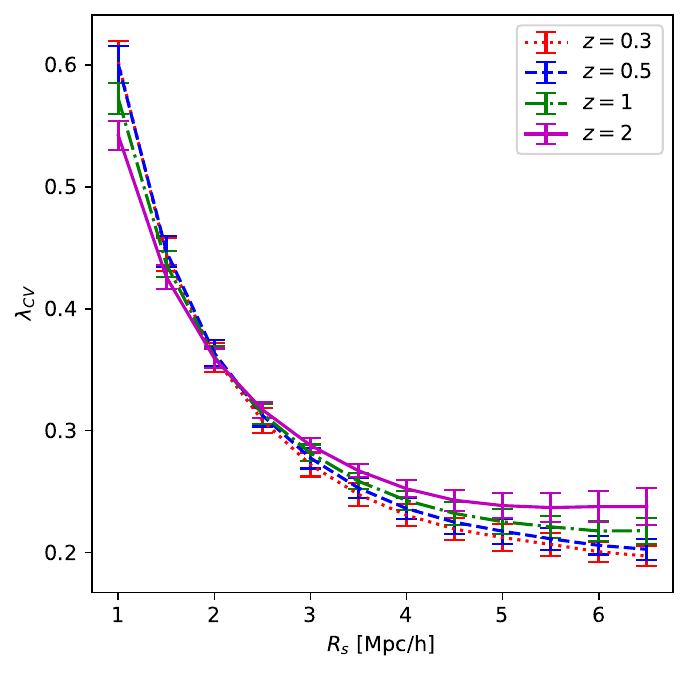}
  \caption{The mean of $\lambda_{CV}$ for different redshifts, as a function of the smoothing length $R_{s}$. The errorbars indicates the $\sigma$ region of set.
          }
     \label{thresh_cRs}
\end{figure}

By repeating the entire process described in the previous subsection for all simulations in the set, Fig.~\ref{thresh_z} presents the mean $\lambda_{CV}$ as a solid line across different redshifts for a specific $R_{s}$. The shaded regions with fainter colors represent the standard deviation $\sigma$ for the set of simulations. Note that the mean value of $\lambda_{CV}$ shows slight variation with redshift. However, it exhibits a clear dependence on the $R_{s}$ used to compute the V-web eigenvalues.
Fig.~\ref{thresh_cRs} shows the mean value of $\lambda_{CV}$ as a function of $R_{s}$, highlighting its decrease with increasing smoothing length. For a large value of $R_{s}$, the eigenvalues tend to have a Gaussian distribution more centered on 0, so $\lambda_{CV}$ tends to the same central value (\ie $\lambda_{CV}\rightarrow 0$). We have not considered $R_{s}$ larger than 6.5 Mpc/h since from this value the voids satisfy $f_v(\overline{\lambda}_{CV}) > 0.99$, thus leaving the value of $\overline{\lambda}_{CV}$ out of the delimited intermediate regime. On the other hand, the resolution of the grid limits the smoothing length to:

 \begin{equation}
    R_{s} \geq \frac{2L}{N_g} \simeq 1 \, \rm{Mpc/h}
\label{eq-lowlim}
\end{equation}
where $L$ is the box size and $N_g$ is the number of cells per grid dimension.

Just as the volume fractions are not exactly conserved because $S(z,\lambda_{CV})$ is never exactly $0$, neither is the value of $\lambda_{CV}$ exactly the same for different redshifts. This is more evident from Fig.~\ref{thresh_z} for $R_{s} = 1$ Mpc/h, where $\lambda_{CV}$ seems to tend to smaller values for higher $z$. On the contrary, for higher values of $R_{s}$ the constant volume threshold appears to increase slightly with $z$. Details on the accuracy of the volume fraction conservation as a function of $z$ and $R_{s}$ are given in Appendix \ref{sec-app_accuracy}. In the next section, we examine in detail how this approximate conservation influences the CW classification and volume fractions across different scales and redshifts. 

We have performed additional analyses in T-web, which have not been explicitly shown here. Our first impression was that in T-web a special threshold also exists, but the conservation of volume fractions is less accurate than in the V-web. However, a more extensive and rigorous analysis is needed to draw strong conclusions, something we leave for possible future work.

\section{Cosmic web at constant volume fraction}
\label{sec-VF_and_MF}

Taking into account the evolution of the CW enables the selection of a specific threshold to be used with V-web for analyzing the spatial distribution of each element. It is crucial to assess whether this classification is consistent with the fundamental properties expected in the CW. 

\subsection{The evolution of the cosmic web at $\lambda_{CV}$}

\begin{figure*}
\centering
\includegraphics[width=17.5cm]{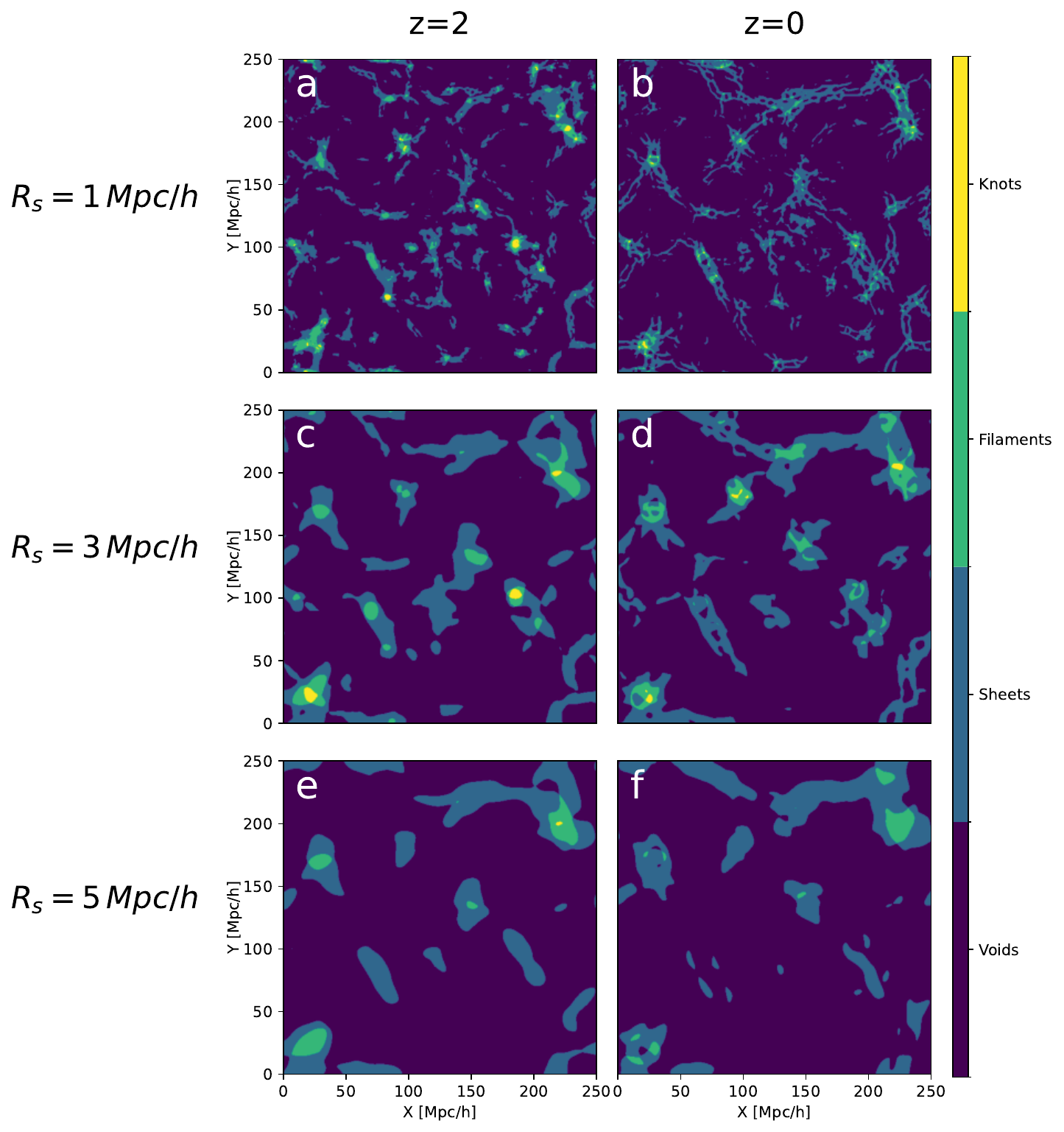}
\caption{V-web classification of the simulation shown in Fig.~\ref{one_page_first}, applying in this case the threshold $\overline{\lambda}_{CV}(z)$ for three different $R_{s}$. As before, the left and right column show $z=2$ and $z=0$ respectively.}
          \label{one_page_last}%
\end{figure*}

A first intuitive approach is to visually inspect how V-web classifies the CW elements when the threshold is $\overline{\lambda}_{CV}$. Fig.~\ref{one_page_last} shows this classification for the same simulation as shown in Fig.~\ref{one_page_first}. This time, we have shown the case for constant volume threshold in three different smoothing lengths. The visual effect of using different $R_{s}$ is the level of small scale detail we can achieve. For all three cases we see a clear evolution of the CW components, from more homogeneous structures at $z=2$ to formations with more complexity at $z=0$. Note that, although each region naturally changes type over time, the use of $\lambda_{CV}$ ensures that the volume fraction remains nearly constant.

In the particular case of $R_{s}=3$ Mpc/h, it is remarkable how the sheets subdivide with time, going from a homogeneous distribution at $z=2$, to increasing number of connections at the present time. Also, as for the filaments, it seems that with time new connections are generated between different knots that allow the exchange of matter between more regions. This is repeated for smaller scales at $R_{s} = 1$ Mpc/h, and for larger scales at $R_{s} = 5$ Mpc/h. In some cases, in the three smoothing lengths the knots and filaments seem to appear and disappear erratically between one time or another. However, this can be explained by the fact that due to their small size they are sensitive to peculiar movements, which can cause them to be lost or included in the two-dimensional slice we use to represent the CW. This effect has to be taken into account also for the other elements, thus limiting the ability of the 2D projections to diagnose the evolutionary behavior. In the following sections we deal with global quantities of the simulations such as volume fraction (section \ref{sec-VF}) or mass fraction (Appendix \ref{sec-MF}).

\subsection{Volume fractions at $\lambda_{CV}$}
\label{sec-VF}
\begin{figure*}
\centering
\includegraphics[width=15cm]{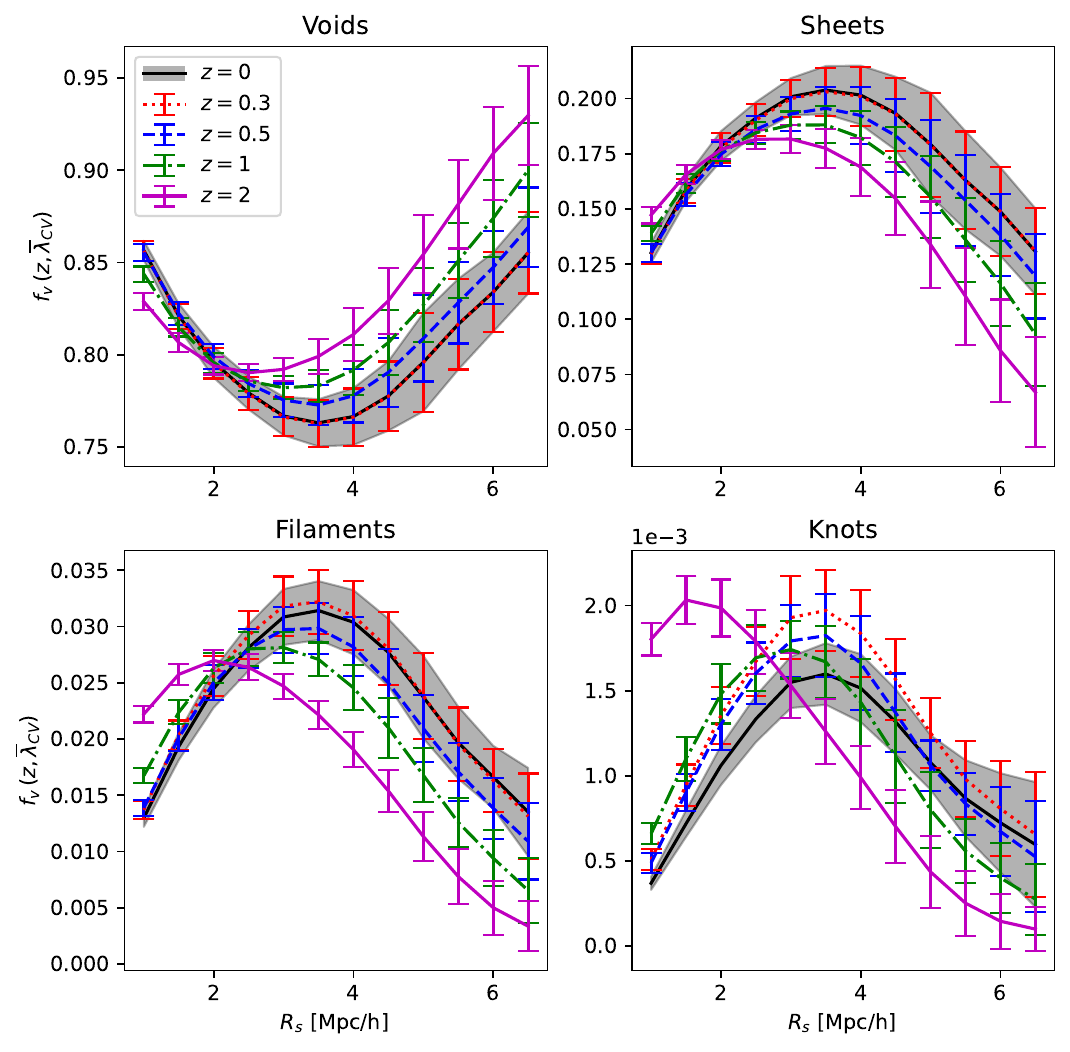}
\caption{Mean volume fractions for different CW elements computed using $\overline{\lambda}_{CV}$ as threshold in the V-web in the $10$ $\Lambda$CDM simulations in which $\lambda_{CV}$ was obtained in section~\ref{sec-evolution}. The value for $z=0$ has been extrapolated and computed using the $\lambda_{CV}$ found for the case $z=0.3$.}
          \label{VF_cRs}%
\end{figure*}

The volume fractions calculated for Fig.~\ref{one_page_last} using the constant volume threshold $\overline{\lambda}_{CV}$ are listed in Table~\ref{table3}. Note the precise conservation of volume fractions for the case $R_{s} = 3$ Mpc/h between $z=2$ and $z=0$ (satisfied to three significant digits in voids, sheets and knots). For the values $R_{s} = 1$ Mpc/h and $R_{s} = 5$ Mpc/h the volume fractions vary slightly from one redshift to another, especially for elements with less statistical presence such as filaments and knots. From Table~\ref{table3} it can also be seen that the volume fractions vary with respect to the $R_{s}$ used at the same $z$. 
\begin{table}
\caption{Constant volume threshold, standard deviation of divergence field and volume fractions computed for the simulation shown in Fig.~\ref{one_page_last}. The values are obtained using the procedure described in section \ref{sec-3.2} for snapshot $z=2$.}             
\label{table3}      
\centering                          
\begin{tabular}{c c c c}        
  & &  $z=2$ &  $z=0$  \\ 
\hline  \\ [-2ex]

 \multirow{5}{7em}{$R_{s} = 1$ Mpc/h} & 
 $\overline{\lambda}_{CV}$ &  $0.20$ &  $0.56$  \\
  & $\sigma_{\theta}$ & $0.21$ & $0.59$ \\
 & Void & 0.83 & 0.84  \\      
    & Sheet & 0.14 & 0.15  \\
    & Filament & 0.021 & 0.016 \\
    & Knot & 0.0017 & 5.1$\times10^{-4}$  \\
\hline \\ [-2ex]                 
\multirow{5}{7em}{$R_{s} = 3$ Mpc/h} & $\overline{\lambda}_{CV}$ &  $0.10$ &  $0.29$  \\
& $\sigma_{\theta}$ & $0.12$ & $0.35$ \\
 & Void & 0.79 & 0.79  \\      
    & Sheet & 0.19 & 0.19  \\
    & Filament & 0.025 & 0.025 \\
    & Knot & 0.0013 & 0.0013  \\
\hline \\ [-2ex]                 
\multirow{5}{7em}{$R_{s} = 5$ Mpc/h} & $\overline{\lambda}_{CV}$ &  $0.078$ &  $0.23$  \\
& $\sigma_{\theta}$ & $0.088$ & $0.26$ \\
 & Void & 0.83 & 0.83  \\      
    & Sheet & 0.16 & 0.15  \\
    & Filament & 0.012 & 0.015 \\
    & Knot & 4.5$\times10^{-4}$ & 4.8$\times10^{-4}$  \\
\hline    
\end{tabular}
\end{table}
The analysis for the set of $10$ simulations is presented in Fig.~\ref{VF_cRs}, showing the volume fractions calculated in the constant volume threshold (we leave the discussion of mass fractions for Appendix \ref{sec-MF}). For $R_{s}$ less than $ \sim 3$ Mpc/h, we see that void volume fraction and $R_{s}$ are inversely correlated. At larger values of $R_{s}$, the correlation is reversed and the presence of voids is bigger for larger smoothing. This trend seems to be inverted for the other elements

We can understand this behavior by looking at the dominant element. At first glance, Fig.~\ref{one_page_last} shows that below $R_{s}$ $ \sim 3$ Mpc/h the amount of space occupied by voids increases as $R_{s}$ decreases. This is because, as we increase the complexity of the structure, the filaments, sheets and knots are subdivided into finer and smaller parts separated from each other by voids. If we assume a $R_{s}$ tending to $0$, the field structure disappear except for the small regions of DM haloes which are stationary under eigenvalue formalism. In this case we have dominance of voids since all haloes/nodes are contracted effectively to volume-less points. If we increase in $R_{s}$ above $ \sim 3$ Mpc/h the voids tend to increase, but for another reason. As the variance of the Gaussian window function increases, the universe tends to become more and more homogeneous and inhomogeneities tend to be smaller. Since the perturbations are where the sheets, filaments and knots originate, this implies an increase of voids against them. 

Note that in Fig.~\ref{VF_cRs} the case for $z=0$ has been included, but in principle it is not possible to define $\overline{\lambda}_{CV}(z=0)$ since we need to use a non-zero redshift in the function $S(z,\lambda_{th})$. However, as $\lambda_{CV}$ shows a slight variation with time, we can consider that $\overline{\lambda}_{CV}$ for a $z$ close to $0$ can be valid also for $z=0$, thus drawing the volume fraction for this redshift from the $\overline{\lambda}_{CV}(z=0.3)$. The results in this extrapolation have very similar values for voids and sheets, although they differs for more sensitive elements such as filaments and knots. Considering all redshifts, the volume fractions are observed to be of a similar order of magnitude, although they exhibit slight variations across different $z$ values. This is due to the fact that volume conservation is not perfect, and $\lambda_{CV}$ is not strictly constant over time (see Appendix \ref{sec-app_accuracy}). In fact, the regime where $\lambda_{CV}$ is most stable over time (between $2$ and $4$ Mpc/h depending on the element) is also the regime in which appears the best volume conservation, there being a crossover of all lines at those values of $R_{s}$. This effect indicates that both volume fraction and $\lambda_{CV}$ conservation tend to occur more accurately over a range of intermediate values of the smoothing length. Since the Gaussian and top-hat smoothing windows also have no fixed choice in the standard analysis, the logic of the constant volume threshold may also help to resolve this ambiguity.

If we compare the volume fractions from the V-web using $\lambda_{CV}$ with those obtained by other widely used techniques, such as Nexus+ (\cite{Cautun2014}), we get very similar results. For $z=0$, the mean fractions for smoothing lengths where conservation is more accurate ($R_{s} = 3$ Mpc/h) are [0.77, 0.20, 0.031,0.0015] for voids, sheets, filaments and knots respectively. For Nexus+, these values are [0.77,0.18,0.06,<0.001], with both methodologies differing mainly in the most sensitive elements such as filaments and knots.

\section{Application of the $\lambda_{CV}$ threshold to generic N-body simulations}
\label{sec-application}

The objective of this section is to provide a generic fit of $\lambda_{CV}$ that can be used in different types of cosmological simulations that are not necessarily restricted to the characteristics of the set used in this work. The idea is to give the community an easy way to obtain the value of $\lambda_{CV}(z,R_{s})$ for a generic cosmological simulation based only on the redshift and Gaussian smoothing scale specified. An example application is presented in section \ref{sec-B800}, applying the fitting function obtained as the V-web threshold in a full N-body simulation with different initial conditions, box size, number of particles and cosmology than those used in the main set. 

\subsection{Approximating the dependence on scale and redshift}
\label{sec-fit}
Given the reduced redshift dependence of $\lambda_{CV}$, we approximate  $\lambda_{CV}^{\rm fit}$ for each $R_{s}$ to a line of slope $a$ and vertical intercept $b$:

\begin{equation}
    \lambda_{CV}^{\rm fit} = az + b
\label{eq-fit}
\end{equation}

where $a$ and $b$ depend on the $R_{s}$ used. The function $a$ quantifies the deviation from the constant value in $z$, being negative for small $R_{s}$ and positive for high $R_{s}$. Function $b$ contain the general behavior of $\lambda_{CV}$ as a function of scale. Fig.~\ref{plot-fit} shows the values of $a$ (left panel) and $b$ (right panel) measured for the main set of simulations, as well as the non-linear least squares fit applied. An exponential function can be used to approximate the dependence of both parameters on $R_{s}$, so the fit equations result in:

\begin{figure*}
\centering
\includegraphics[width=15cm]{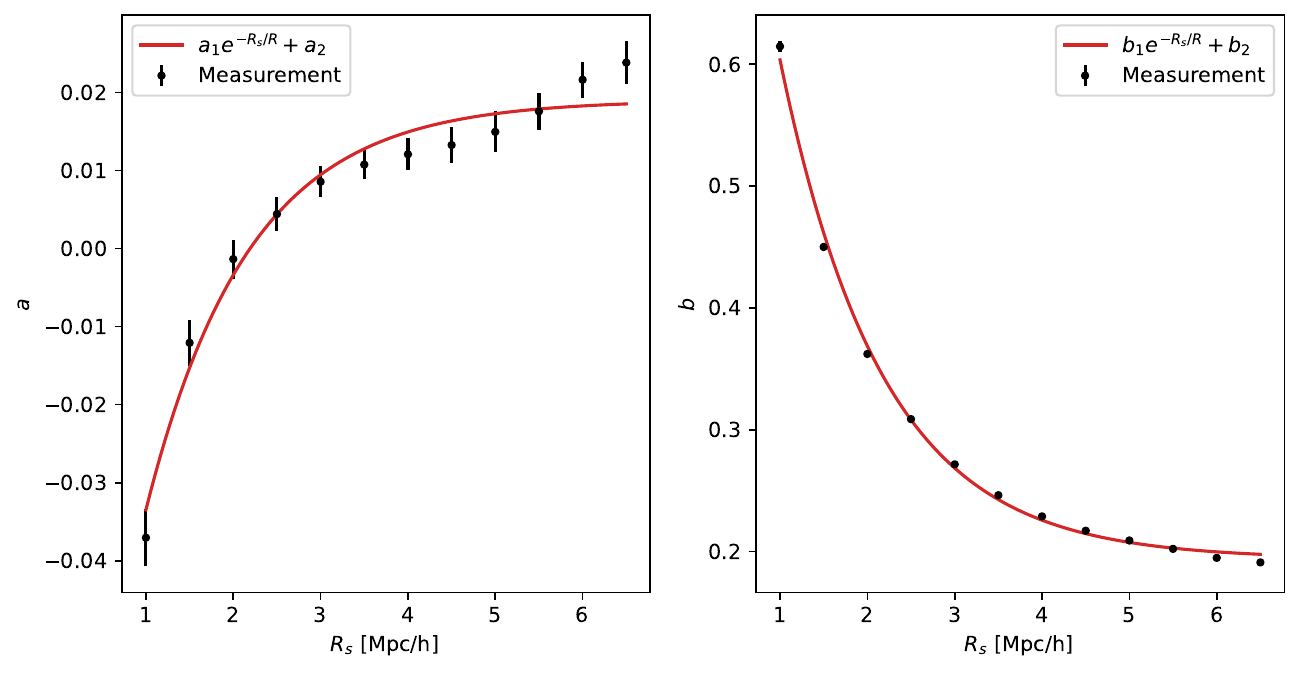}
\caption{The parameters $a$ and $b$ defined in Eq.~(\ref{eq-fit}) and measured by least squares applied to each value of $R_{s}$ used to obtain $\lambda_{CV}$ in the MG-COLA set of simulations. The error bars (small in the right panel) are given by the square root of the diagonal elements in the covariance matrix. Both parameters have been approximated to an exponential type function plotted in red, whose constants are given in Table~\ref{tab-fit}. The result of the approximation is a fit $\lambda_{CV}^{\rm fit}(R_{s},z)$ represented in Fig.~\ref{thresh_z} by dashed lines.}
\label{plot-fit}
\end{figure*}
\begin{equation}
    a(R_{s}) = a_1 e^{-R_{s}/R} + a_2\,\,,
\label{eq-fit2}
\end{equation}
\begin{equation}
    b(R_{s}) = b_1e^{-R_{s}/R} + b_2\,\,,
\label{eq-fit3}
\end{equation}

with the parameters values obtained from our set and given in Table~\ref{tab-fit}. The standard deviation errors of the fit parameters are calculated by the square root of the diagonal of the covariance matrix. $\lambda_{CV}^{\rm fit}$ is represented by dashed lines in Fig.~\ref{thresh_z}, confirming that the fit falls within the $\sigma$ cosmic variance at all represented $R_{s}$ values. The procedure for using the fit is as follows: first, substitute in Eqs.~(\ref{eq-fit2}) and (\ref{eq-fit3}) the $R_s$ used, and input the redshift into Eq.~(\ref{eq-fit}) to get $\lambda_{CV}^{\rm fit}$. Next, since $\lambda_{CV}^{\rm fit}$ has the role of $\lambda_{th}$ in Eq.~(\ref{eq-norm}), compute $\sigma_{\theta}^0$ and $\sigma_{\theta}(z)$ to obtain $\overline{\lambda}_{CV}(z)$. This is the threshold to be used in the V-web for the snapshot at redshift $z$. Note that if $z=0$ then the second step is not necessary.

For example, to determine the threshold for $R_s = 4$ Mpc/h and $z=0$ it is sufficient to calculate $b(R_s)$ from Eq.~(\ref{eq-fit3}), yielding $\lambda_{CV}^{\rm fit}=0.23$. If $z=2$ is considered instead for same $R_s$, the procedure requires calculating $a(R_s)$ from Eq.~(\ref{eq-fit2}), followed by applying the re-scaling (\ref{eq-norm}) to the resulting $\lambda_{CV}^{\rm fit}$. Applying this approach to the simulation from our set shown in Fig.~\ref{one_page_first}, where $\sigma_{\theta}^0 = 0.30$ and $\sigma_{\theta}(z=2) = 0.10$ for $R_s = 4$ Mpc/h, the resulting threshold for V-web at snapshot $z=2$ is $\overline{\lambda}_{CV}^{\rm fit} = 0.085$. 

\begin{table}
\caption{Parameters used in Eq.~(\ref{eq-fit2})}             
\label{tab-fit}      
\centering                          
\begin{tabular}{c c c}        
    & Value & Error\\ 
\hline \\ [-2ex]
$a_1$ & $-0.123$ & $\pm 0.009$ \\ 
$a_2$ & $0.019$ & $\pm 0.001$ \\ 
$b_1$ & $0.960$ & $\pm 0.018$ \\ 
$b_2$ & $0.194$ & $\pm 0.001$ \\ 
$R$ & $1.174$ Mpc/h & $\pm 0.017$ Mpc/h\\ 
\end{tabular}
\end{table}

\subsection{$\lambda_{CV}$ in a N-body simulation outside the set}
\label{sec-B800}

We have reproduced the whole process of $\lambda_{CV}$ calculation for an example full N-body simulation. Since the normalization of Eq.~(\ref{eq-pe}) deals with $\sigma_{\theta}$ ratios, the dependence of $\lambda_{CV}$ on similar cosmological parameters is very small and its value can be approximated by $\lambda_{CV}^{\rm fit}$ for off-set $\Lambda$CDM simulations.

\begin{figure*}
\centering
\includegraphics[width=15cm]{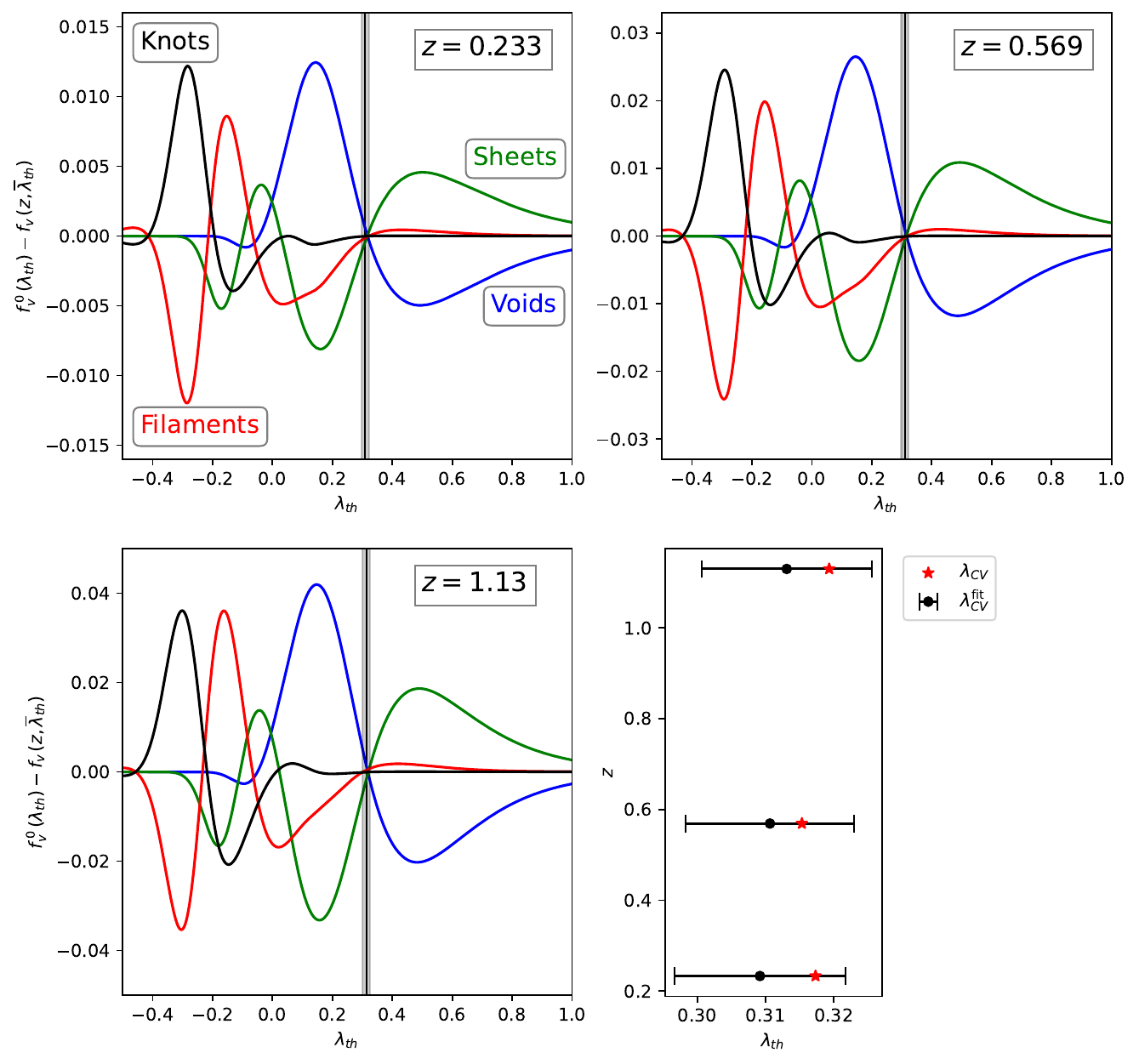}
\caption{The first three panels show the differences in volume fractions at different redshifts, calculated for an off-set Gadget simulation. The constant volume threshold phenomenon appears for all classified elements despite the fact that the cosmology, box size, initial conditions and code are different from those of the MG-COLA set used in the rest of the work. The black vertical area shows the calculated range for $\lambda_{CV}^{\rm fit} \pm 3\sigma_{\rm fit}$ with smoothing length $R_{s}=2.5$ Mpc/h used in this example. The last panel explicitly shows the measurement $\lambda_{CV}$ and fit $\lambda_{CV}^{\rm fit}$ values.}
\label{plot-fitB800}
\end{figure*}

As a working example N-body simulation we take a run done with Gadget-2 code \citep{Springel2005}
with a use of $N_p=1280^3$ particles each representing $m_p=1.8\times 10^{10} M_{\sun}/h$ matter element,
placed in a periodic box of $L=800$Mpc/h \citep{Hellwing2017}. The cosmological parameters used to generate the initial conditions and fixed the expansion history were that of the 7-year data release of the WMAP mission \citep{WMAP7}. Fig.~\ref{plot-fitB800} shows the volume fraction differences for the Gadget simulation in a similar way as drawn in Fig.~\ref{cRs3_Difvsth} for a simulation from the MG-COLA set. In the first three panels a particular redshift is depicted, and for each the corresponding $\lambda_{CV}^{\rm fit}$ is represented by the vertical black line (whose thickness matches  three times the propagated uncertainty $\sigma_{ \rm fit}$). It can be seen that the volume fractions of all elements remain constant at a specific threshold value, confirming the existence of the constant volume threshold in the off-set simulation. The right panel shows with red stars the value of $\lambda_{CV}$ calculated for the N-body simulation using the standard procedure of minimizing the $S(z,\lambda_{th})$ function. This falls within the values expected by $\lambda_{CV}^{\rm fit}$ for the represented redshifts.

\section{Concluding remarks}
\label{sec-conclusions}

The study of the cosmic web is essential for a comprehensive understanding of large-scale structure, so its analysis requires a consistent classification based upon solid physical principles. Among the various existing techniques for identifying the structures that make up the CW, Hessian matrix-based methods stand out for their simplicity, theoretical motivation, and potential applicability to observations. These methods allow the identification of voids, sheets, filaments, and knots, depending on the principal axes of collapse of the gravitational potential (T-web) or the velocity field (V-web). However, both classifications rely on a threshold $\lambda_{th}$ that has not been assigned with a physical value and has so far been chosen primarily through visual inspection.

In this work, we address this problem for the case of the V-web classification by using the statistical principles governing Gaussian eigenvalues (\eg Doroshkevich's formalism), focusing our analysis on the probability of each CW element’s presence quantified by its volume fraction. To analyze the non-linear evolution we use a set of 10 $\Lambda$CDM simulations generated with MG-COLA under different initial conditions. In this set, the V-web method is applied using a threshold normalized by the standard deviation of the velocity divergence field. As shown in Fig.~\ref{cRs3_Difvsth}, this approach reveals a unique positive threshold at which all volume fractions remain constant over time. Figs.~\ref{thresh_z} and \ref{thresh_cRs} indicate that this constant volume threshold $\lambda_{CV}$ depends on scale but shows little variation with cosmic variance or redshift. The resulting classification (Fig.~\ref{one_page_first}) agrees with the choice by visual inspection reported in previous works, such as \cite{Hoffman2012}. Moreover, the volume fractions obtained using $\lambda_{CV}$ at scales close to $R_s = 3$ Mpc/h, where conservation is most precise, are consistent with those predicted by other methods, such as Nexus+ \citep{Cautun2014}. Similarly, the evolution of the mass distribution across different CW elements aligns with previous findings. 

As an alternative to the arbitrary threshold in the V-web method, we provide a fit for $\lambda_{CV}$. This fit depends on the applied Gussian smoothing length $R_{s}$ and the redshift $z$ in which V-web is used. If $z=0$, then the threshold $\lambda_{CV}^{\rm fit}$ that keeps the volume constant is equal to the function $b(R_{s})$ of Eq.~\ref{eq-fit3}. If $z>0$ then one must also calculate the function $a(R_{s})$ of Eq.~\ref{eq-fit2} and substitute both in Eq.~\ref{eq-fit} to obtain $\lambda_{CV}^{\rm fit}$, to finally calculate the V-web threshold by means of the re-scaling defined by Eq.~\ref{eq-norm}. Finally, we test the universality of this fit using a Gadget simulation with characteristics distinct from those of the MG-COLA set. As shown in Fig.~\ref{plot-fitB800}, the simulation outside the set demonstrates volume fraction conservation that aligns with $\lambda_{CV}^{\rm fit}$.

The clear manifestation of $\lambda_{CV}$ motivates an universal threshold for the V-web, that is both physically motivated and independent of user choice. Our results hints also towards a new  fundamental physical behavior in this type of Hessian classification.
This work aims to open the door to a new perspective with which to understand the large-scale structure, and to take a step in improving our definition of the cosmic web.

\begin{acknowledgements}
The authors are grateful for inspiring discussion and comments form Adi Nusser and Mariana Bravo Jaber, received at early stage of this work. EO thanks Yehuda Hoffman for his advice and Anastasia Hrabarchuk for her insistence. EO and AK are supported by the Ministerio de Ciencia e Innovaci\'{o}n (MICINN) under research grant PID2021-122603NB-C21. EO received predoctoral fellowship from MICINN (FPI program, Ref. PRE2022-102254). AK further thanks Portishead for dummy. WAH acknowledges the support from the research projects funded by the National Science Center, Poland, under agreement number \#2018/30/E/ST9/00698 and \#2018/31/G/ST9/03388.

\end{acknowledgements}

%
%

\bibliographystyle{aa} 
\bibliography{biblio} 

\begin{appendix} 
\section{Volume fraction differences in $\lambda_{CV}$}
\label{sec-app_accuracy}

\begin{figure}
\centering
\includegraphics[width=\hsize]{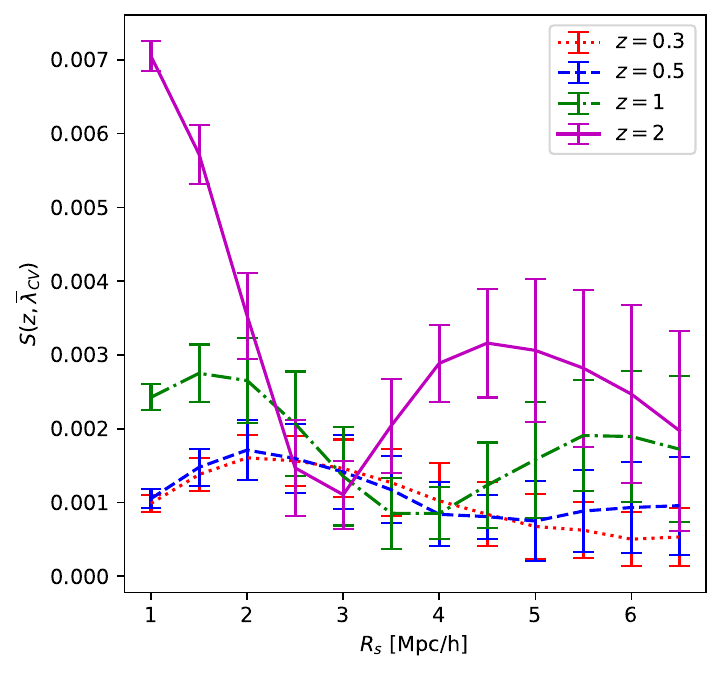}
  \caption{The function S defined in Eq.~(\ref{S}), evaluated at the constant volume threshold $\lambda_{CV}$, for different redshifts and comoving smoothing lengths $R_{s}$. The errorbars indicates the $\sigma$ region of the cosmic variance.}
     \label{S_cRs}
\end{figure}

To understand why the volume fractions may not be exactly conserved, let us first assume that it is in fact conserved for a certain constant $\lambda_{CV}$. Suppose $z_1$ and $z_2$ arbitrary redshifts different from zero, for which we have obtained a constant volume threshold $\lambda_{CV}^1$ and $\lambda_{CV}^2$ respectively, thus satisfying $S(z_1,\lambda_{CV}^1)=0$ and $S(z_2,\lambda_{CV}^2)=0$. Since we assume that conservation is satisfied for all elements one by one, then

\begin{equation}
    f_{v}^{0,\rm E}(\lambda_{CV}^{1}) - f_{v}^{\rm E}(z_{1},\overline{\lambda}_{CV}^{1}) = 0
\label{consz1}
\end{equation}

\begin{equation}
    f_{v}^{0,\rm E}(\lambda_{CV}^{2}) - f_{v}^{\rm E}(z_{2},\overline{\lambda}_{CV}^{2}) = 0
\label{consz2}
\end{equation}

for any CW element E. If the constant volume threshold is equal for all redshifts, then the difference of the above equations is 

\begin{equation}
    f_{v}^{\rm E}(z_{1},\overline{\lambda}_{CV}^{1}) - f_{v}^{\rm E}(z_{2},\overline{\lambda}_{CV}^{2}) = 0
\label{consz1z2}
\end{equation}

with the volume fractions being equal over all time. Eq. (\ref{consz1z2}) may not be satisfied for two reasons. The first is that the one-to-one conservation of Eqs.~(\ref{consz1}) and (\ref{consz2}) is not exact, so also in general $S(z,\lambda_{CV})\neq0$. Fig.~\ref{S_cRs} represents this effect in the set of MG-COLA simulations by the value of $S(z,\lambda_{CV})$ at different smoothing lengths. Note that for higher redshifts $S(z,\lambda_{CV})$ is closer to 0 at intermediate smoothing lengths near $R_{s} = 3$ Mpc/h.

\begin{figure}
\centering
\includegraphics[width=\hsize]{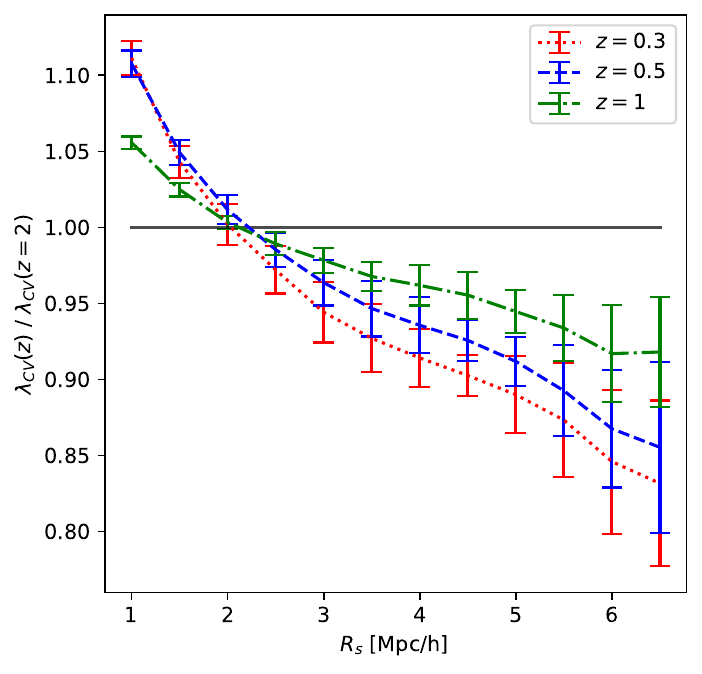}
  \caption{Ratio between $\lambda_{CV}$ calculated for a snapshot with redshift $z$, and the $\lambda_{CV}$ calculated for a snapshot with redshift $z=2$. The black line shows the value $1$.}
     \label{th_thz2}
\end{figure}

The second reason why Eq.~\ref{consz1z2} may not be fulfilled is that $\lambda_{CV}^{1}\neq\lambda_{CV}^{2}$, so the first terms of Eq.~\ref{consz1} and Eq.~\ref{consz2} would not cancel. This cause that even if we had exact pairwise cancellation, it would not be fulfilled among all redshifts. As shown in Fig.~\ref{thresh_z}, the different $\lambda_{CV}$ are not always exactly constant, accentuating this effect for small and large $R_{s}$. Fig.~\ref{th_thz2} represents the ratio between different $\lambda_{CV}(z)$, where we have arbitrarily chosen $z=2$ as a reference. It can be noticed more clearly how this fraction moves away from unity, again for extreme $R_{s}$ and more markedly for $z$ farther away from the reference value. In the case of $R_{s}$ smaller than $2$ Mpc/h $\lambda_{CV}$ grows for small $z$, while for $R_{s}$ larger than $2$ Mpc/h it grows for large $z$, as can also be seen in Fig.~\ref{thresh_z}. 

The combination of both effects produces the differences in $f_{v}(z,\overline{\lambda}_{CV})$ between redshifts in Fig.~\ref{VF_cRs}, complying for all elements the existence of a narrow regime of $R_{s}$ between $2$ and $3$ Mpc/h in which the volume fraction is constant in a more precise manner.

\section{Density of the Cosmic Web}
\label{sec-MF}
In the same way that we have used the volume fraction as a measure of the statistical presence, the mass fraction gives us clues about the evolution of the dark matter distribution in CW. More specifically, measuring the time dependence gives us information on the mass transfer between the different elements, as well as which of them are more populated with matter than others. We can define the mass fraction of a CW element as the total mass that is classified as that element divided by the total mass inside the simulated box. If we divide the mass fraction $f_{m}$ by the volume fraction $f_{v}$, we get an estimate of the total density $\Delta$ occupied by a certain element:

\begin{equation}
    \Delta (z,\lambda_{th}) = \frac{f_{m}(z,\lambda_{th})}{f_{v}(z,\lambda_{th})}
\label{eq-dens}
\end{equation}

\begin{figure*}
\centering
\includegraphics[width=15cm]{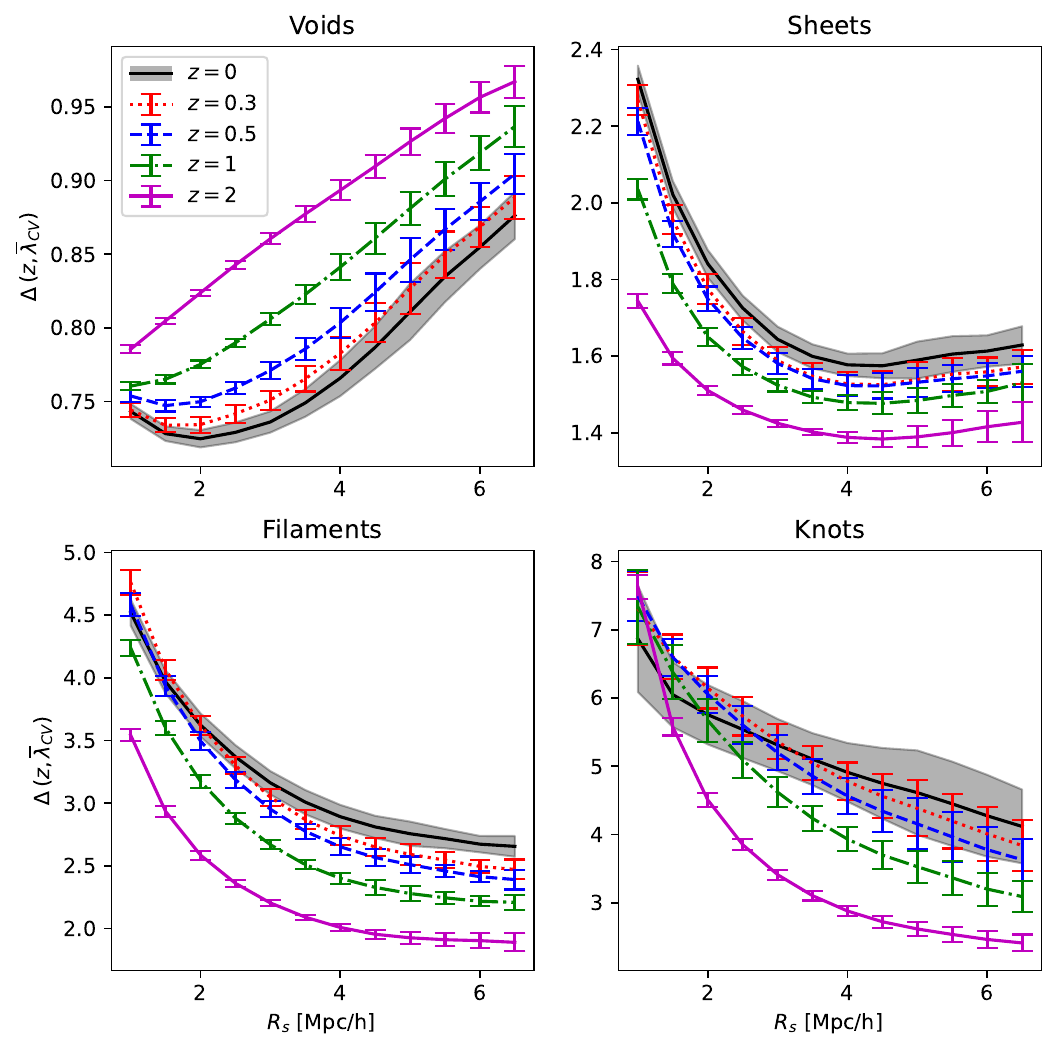}
\caption{Total density of each CW element, defined as mass fraction divided by volume fraction computed using $\overline{\lambda}_{CV}$.}
          \label{Dens_cRs}%
\end{figure*}

Fig.~\ref{Dens_cRs} shows the density of each of the CW components at constant volume threshold. When applying the methodology of section~\ref{sec-evolution} to mass fractions instead of volume fractions there is no conservation at any threshold. This causes the density to have a clear evolution as can be seen in Fig.~\ref{Dens_cRs}, including the extrapolation to $z=0$. As $\Delta$ decreases with time for voids in all $R_{s}$ values, the opposite occurs for sheets and filaments. Similar behavior has also been observed in other works on mass evolution in the CW \citep[\eg][]{Cautun2014,Zhu2017}.

\end{appendix}

\end{document}